\documentclass[sn-mathphys,Numbered]{sn-jnl}
\usepackage{graphicx}%
\usepackage{multirow}%
\usepackage{amsmath,amssymb,amsfonts}%
\usepackage{amsthm}%
\usepackage{mathrsfs}%
\usepackage[title]{appendix}%
\usepackage{xcolor}%
\usepackage{textcomp}%
\usepackage{manyfoot}%
\usepackage{booktabs}%
\usepackage{algorithm}%
\usepackage{algorithmicx}%
\usepackage{algpseudocode}%
\usepackage{listings}%
\usepackage{amssymb}
\usepackage{centernot}
\usepackage{lipsum}
\usepackage[font=footnotesize,labelfont=bf]{caption}
\usepackage{subfigure}
\usepackage{float}
\usepackage{amsmath}
\usepackage{fancyhdr}
\usepackage{blindtext}

\theoremstyle{thmstyleone}%

\theoremstyle{thmstyletwo}%

\theoremstyle{thmstylethree}%

\begin{document}
\title{The noncommutative phase space coordinates in Moyal  framework:  $2$-dimensional quantum  harmonic oscillator and Painlev\'e second equation}
\author[1]{Irfan Mahmood \thanks{ Corresponding author:irfan.chep@pu.edu.pk}}
\author[2,3,4]{Wen-Xiu Ma \thanks{Corresponding author: wma3@usf.edu}}
\affil[1]{Center for High 
 Energy Physics, University of the Punjab, Pakistan}

 \affil[2]{Department of Mathematics, Zhejiang Normal University, Jinhua, 321004, Zhejiang, China }
\affil[3]{Department of Mathematics and Statistics, University of South Florida, Tampa, FL 33620-5700, USA}
\affil[4]{Material Science Innovation and Modelling, North-West University, Mafikeng Campus, Mmabatho, 2735, South Africa}

\affil{Corresponding Author: Irfan Mahmood (mahirfan@yahoo.com ) }
\abstract{
This work aims to explore the  application of Moyal noncommutative formalism with introduction of  new phase space coordinates transformation inculdign  multi components antisymmetric tensors. These transformations  connect the quantum phase space  coordinates to Purely noncommuting phase space coordinates which are consistent with the Moyal noncommutative brackets.  We discuss the integrability of $2$-dimensional quantum harmonic oscillator and Painlevé second equation with their noncommutative structures through the implementations of  noncommutative phase space coordinate transformations.}

\maketitle
\section{Noncommutativity: Moyal product}
The noncommutative (NC) extension of field theories   is a   quite interesting research direction in modern  theory of integrable systems for  exploring their NC Hamiltonian structures with associated  symmetries.  These extensions are acknowledged as the quantum  generalization of classical field theories with the addition of space-space commutativity which  are reducible to their classical and quantum  analogs under respective limits. The idea of NC extension under the star product, initially conceived  in early  work  of H.J. Groenewold  \cite{HJ} and subsequently was improved by  J.E. Moyal \cite{JE}  which is known as  Moyal star product. This  formalism incorporates  space-space and space-momentum noncommutativity rather then taking Heisenberg quantum commutation relations for the space-momentum  coordinates. In this context,  NC analogs of various integrable systems  and  field equations  have been investigated in  Moyal formalism  \cite{HT} - \cite{MB} and in other related frameworks \cite{NHO} - \cite{PAD2} to reveal physical understanding wit their noncommutative structures. The  Moyal-product found very effective to construct the NC analogs of classical theories, which can be realized as  
deformed theories,  from the commutative ones.   
In an explicit way the Moyal product  (NC product) of arbitrary objects $ f(x)$  and $ g (x)$ is  defined by
\begin{eqnarray}\label{1}
f(x) \star g(x)=exp(\frac{i}{2}\theta^{\mu\nu} \frac{\partial}{\partial x^{'\mu}}\frac{\partial}{\partial x^{''\nu}})f(x^{'})g(x^{''})_{x=x^{'}=x^{''}},
\end{eqnarray} 
and its truncated form upto first order may be written as below 
\begin{eqnarray}\label{2}
f(x) \star g(x)=f(x)g(x)+\frac{i}{2}\theta^{\mu\nu}\frac{\partial f}{\partial x^{\mu}}\frac{\partial g}{\partial x^{\nu}}+ \mathcal{O}(\theta^2).\end{eqnarray} 
Here $\theta^{\mu \nu}  $ is a parameter which describes the noncommutativity. Another notable point about $\theta^{\mu \nu}  $ in  Moyal product, it is anti-symmetric, Lorentz invariant and describes an infinitesimal NC extension of ordinary product  as its square and higher powers are omitted. Under the
 commutative limit
$\theta^{\mu \nu} \rightarrow 0 $
the above expression will  reduce to the ordinary product as   $ f \star g = f. g$. 
The NC spaces are characterized by the noncommutativity of the  coordinates, if $ x^{\mu} $ are the  coordinates, then the noncommutativity under the Moyal product is calculated as
$ [x^{\mu},x^{\nu}]_{\star}=i\theta^{\mu \nu} $ 
where  parameter $ \theta^{\mu\nu}$ describes noncommutativity  and $ [x^{\mu},x^{\nu}]_{\star}$ is a commutator under the Moyal product. 

This formalism incorporates space-space and space-momentum noncommutativity rather then the Heisenberg quantum commutation relations  only for the corresponding  canonical coordinates. These transformations imply to construct noncommutative analogs of $2$-dimensional quantum  harmonic (QH) oscillator with its superintegrability and the derivation of noncommutative  Painlevé second equation. The associated deformed Hamiltonians of these systems involve some extra terms which reveal a deep physical understanding  about the additional symmetries with noncommunicative structures. The noncommutative phase space coordinates transformation for Painlevé second equation  further implies to generate the first Yablonskii Vorobev polynomial solution of that equation.
\section*{Proposition 1.1}
The Moyal bracket for any two  arbitrary functions $f(x^{\mu},x^{\nu})$ and $ g(x^{\mu},x^{\nu})$ may be connected to the Poisson bracket. \\ 
\textbf{Proof:}
Changing the positions of  $f(x^{\mu},x^{\nu})$ and $g(x^{\mu},x^{\nu})$ in expression \ref{2}  and the subtracting  expression \ref{2} from the resulting expression, we get the following result 
\begin{eqnarray}\label{MP}
f(x^{\mu},x^{\nu}) \star g(x^{\mu},x^{\nu}) -  g(x^{\mu},x^{\nu}) \star f(x^{\mu},x^{\nu})=\frac{i}{2}\theta^{\mu\nu} \frac{\partial f}{\partial x^{\mu}}\frac{\partial g}{\partial x^{\nu}}  - \frac{i}{2}\theta^{\mu\nu}\frac{\partial g}{\partial x^{\mu}}\frac{\partial f}{\partial x^{\nu}}.\end{eqnarray} 
or
\begin{eqnarray}\label{MP}
[f(x^{\mu},x^{\nu}), g(x^{\mu},x^{\nu})]_{Moyal} = \frac{i}{2}\theta^{\mu\nu} \{ f(x^{\mu},x^{\nu}), g(x^{\mu},x^{\nu})\}_{PB}
\end{eqnarray} 
here on left hand side is  Moyal bracket which is connected to the Poisson bracket through $\frac{i}{2}\theta^{\mu\nu} $.
\section*{Remark 1.1}
The noncommutativity among the coordinates under the Moyal product is recognized up to the power $+1$ of the deformation (noncommuting) parameter $\theta^{\mu \nu}$. It means in case of non-self commutator in Moyal framework we may omit the term involves $\theta^{\mu \nu}$ with power $+2$ and higher powers. 
\subsubsection*{Note:} Here to discuss pure noncommutative analogs of $2$-D quantum harmonic oscillator and Painlev\'e second equation, we start with their hamiltonians which include the phase coordinates. Further we construct a set of transformation which connects  noncommuting phase  coordinates $\left(X, Y ,P_{X}, P_{Y}\right)$ to old canonical coordinates $\left(x, y ,p_{x}, p_{y}\right)$. The last section encloses the presentation of noncommuataive Painlev\'e second equation with associated noncommutative phase coordinates transformations which are consistent with Moyal bracket. \\
Let  the NC phase coordinates $ \left(X, Y ,P_{X}, P_{Y}\right)=\left(X^{1}, X^{2}, X^{3}, X^{4}\right) $ satisfy the Moyal product then the Moyal bracket for any two coordinates is defined  as   
\begin{equation}\label{XNC}
[X^\mu, X^\nu] = i\theta^{\mu\nu}.
\end{equation} 
 Let for any two coordinates say for $X^{1}$ and $X^{4}$ we may construct the above relation (\ref{XNC}) under the Moyal product  (\ref{2})
$$ X^{1}*X^{4}= X^{1}.X^{4} + \frac{i}{2}\theta^{14},$$
$$ X^{4}*X^{1}= X^{1}.X^{4} + \frac{i}{2}\theta^{41},$$
 after subtracting above expressions, we get
$$ X^{1}*X^{4}-X^{4}*X^{1}=[X^1, X^4]=   i\theta^{14}.$$
Similarly we can calculate the rest of the commutation relations with a $16$ components of anti-symmetric tensor.
 
\section*{Theorem $1.1$:}
The following noncommutaive phase transformations 
\begin{equation}\label{Tr}
\left\{
\begin{aligned}
& X^{1} = X = x + \frac{\theta^{12}}{\hbar} p_{y}, \\
& X^{2} = Y = y + \frac{\theta^{21}}{\hbar} p_{x}, \\
& X^{3} = P_{X} = p_{x} + \frac{\theta^{23}}{\hbar} p_{y}, \\
& X^{4} = P_{Y} = p_{y} + \frac{\theta^{14}}{\hbar} p_{x}.
\end{aligned}
\right.
\end{equation}
with a $16$ components of anti-symmetric tensor
\begin{align}
 \theta^{\mu \nu}=\left(\begin{array}{cccc}
\theta^{11} & \theta^{12} & \theta^{13} & \theta^{14} \\
-\theta^{12} & \theta^{22} & \theta^{23} & \theta^{24} \\
-\theta^{13} & -\theta^{23} & \theta^{33} &  \theta^{34} \\
-\theta^{14} & -\theta^{24} & -\theta^{34} & \theta^{44}
\end{array}\right). 
\end{align}\
are consistent with Moyal bracket.
\section*{Proof:}
To prove the consistency of phase coordinates transformation (\ref{Tr}) with Moyal bracket, let compute first $ [X^1, X^2]$. From above transformations we have $[X^1, X^2]= \frac{\theta^{21}}{\hbar}[x, p_x]  + \frac{\theta^{12}}{\hbar}[p_y, y]= i\epsilon $ providing $[x_i, p_{x_i}] = i\hbar$ for quantum harmonic oscillator with $ \theta^{12} = \frac{\epsilon}{2}$.  Now we calculate  $[X^1, X^3]= [x,, p_x] + \frac{\theta^{13}}{\hbar}[x,p_y]+\frac{\theta^{12}}{\hbar}[p_y, p_x]+ \frac{\theta^{12}}{\hbar}\frac{\theta^{13}}{\hbar}[p_y, p_y]=  [x,, p_x]= i\hbar= i \theta^{13}$, similarly $ [X^1, X^3]= i \theta^{14}$. We can continue to calculate the remaining brackets for noncommuting phase coordinates by using above transformations.  
\section*{Remark 1.2:  \hspace{0.1cm} Transformations in operator forms}
The above transformation (\ref{Tr}) which connects the old phase  coordinates ($x_i , p_i$)  to fully  NC phase  coordinates can be expressed in operator form with $p_i = i\hbar  \frac{\partial}{\partial x_i}$. This further implies to show that the momentum operators $P_{X} = i\hbar  \frac{\partial}{\partial x} + i \theta^{23} \frac{\partial}{\partial y} $ and $ P_{Y} = i\hbar  \frac{\partial}{\partial y} + i \theta^{14}  \frac{\partial}{\partial x}$ commute in Moyal framework as the usual momentum operators  commute  in quantum mechanics. Further this can be shown that the compatibility of partial derivatives prevail in noncommuataive framework for $ P_{X}$ and $P_{X}$ in operator forms.
\section*{Proposition $1.2$  \hspace{0.1cm}} 
For the convenience we may set NC parameters as $ \theta^{12}=\epsilon$, $ \theta^{13}=\hbar$, $ \theta^{14}=\delta$, $ \theta^{23}=\eta$,  $ \theta^{24}=\hbar$ and $\theta^{34}=0$. Now the expression for the anti-symmetric  tensor may take the following form
\begin{align}
 \theta^{\mu \nu}=\left(\begin{array}{cccc}
0 & \epsilon & \hbar & \delta \\
-\epsilon & 0 & \eta & \hbar \\
-\hbar & -\eta & 0 & 0 \\
-\delta & -\hbar & 0 & 0
\end{array}\right), 
\end{align}
which is consistent with the Moyal bracket for variables $X^\mu$ with $\mu= 1,2, 3, 4$  are described in transformation (\ref{Tr}) involving these NC  parameters. Here the values of anti-symmetric tensor are zero on diagonal positions due to  the self commutation relations  as $[X^i, X^i]= 0$ for $i= 1,2,3,4$ .\\ 
\section{$2$-Dimensional NC Harmonic Oscillator}
This section includes the derivation of $2$-D NC harmonic oscillator,  its Hamiltonian, super-integrability and associated energy forms  under space-space and space-momentum noncommuataivity as is defined in \ref{2}. Its space-space noncommutative case  has been presented  by B. Muthukumar and P. Mitra \cite{NHO} with few aspects  without  taking arbitrary space-momentum noncommutativity which is a main motivation of this work. \\  
The $2$-Dimensional ($2$-D) quantum harmonic oscillator with mass $m$ and angular velocity $\omega$  is described by following Hamiltonian 
\begin{equation}\label{QH}
H = \frac{1}{2m} \sum_{k=1}^{2} \left( p_k p_k + m^2\omega^2 x_k x_k \right),
\end{equation}
here $p_1= p_x$ , $p_2= p_y$, $x_1= x$, $x_2= y$ and subjected to Heisenberg  quantum commutation  relations $ [x_i,p_j]=i\hbar $, if $i=j$ and $ [x_i,p_j]=0 $, if $i \neq j$, the rest of the commutation relations for all variables $x_i$, $p_j$ of  are zero. 
The computation of  $2$-D NC Harmonic oscillator begins from its quantum analog. 
\subsection{Hamiltonian representation of $2$-D NC harmonic oscillator}
The Hamiltonian of  $2$-D NC harmonic oscillator in terms of NC coordinates can be expressed in following form
\begin{equation} \label{NCH}
H(X, Y, P_X, P_Y) 
= \frac{1}{2m} P_X^2 + \frac{1}{2m} P_Y^2 
+ \frac{1}{2} m \omega^2 X^2 + \frac{1}{2} m \omega^2 Y^2.    
\end{equation}
The above Hamiltonian further can be transformed into an expression that includes NC parameters by substituting the values of $X$, $Y$,  $P_X$ and $P_Y$ from (\ref{Tr}) and then after simplification with following settings 
\begin{equation}
\left\{
\begin{aligned}
& \frac{1}{2M} = \left( \frac{1}{2m} + \frac{\delta^2}{2m \hbar^2} + \frac{m \omega^2 \epsilon^2}{2 \hbar^2} \right), \\
& \frac{1}{2} M \Omega^2 = \frac{1}{2} m \omega^2,
\end{aligned}
\right.
\end{equation}
and
\begin{equation}
\left\{
\begin{aligned}
& D = \frac{\hbar^2}{\hbar^2 + \delta^2 }, \\
& B = \frac{1}{2M}  - \frac{M \Omega^2 \epsilon^2}{2 \hbar^2}, \\
& \frac{1}{2m} = D B,
\end{aligned}
\right.
\end{equation}
we obtain  
\begin{align} \label{NCH}
H =  \dfrac{1}{2M} \left( p_x^2  + p_y^2  \right)+ 
 + \frac{1}{2} M \Omega^2 \left( x^2  + y^2  \right)
 + \dfrac{4\eta}{ \hbar} BD p_x p_y  
 + \frac{1}{2} M \Omega^2  \dfrac{ \epsilon}{\hbar} \left( xp_y -yp_x \right).
\end{align}
or more explicitly 
\begin{align} \label{NCH} 
H= \underbrace{\frac{p_x^2+p_y^2}{2M} +\frac12M\Omega^2(x^2+y^2)}_{\text{ordinary 2-D harmonic oscillator}}+\underbrace{\frac{4\eta}{\hbar}BD\,p_xp_y}_{\text{momentum coupling}} +\underbrace{\frac12M\Omega^2\frac{\epsilon}{\hbar}(xp_y-yp_x)}_{\text{angular-momentum}}.\end{align}
This can be shown that with commutative limit as $\theta^{\mu \nu} \longrightarrow 0$, we obtain original Hamiltonian  (\ref{QH}).  We have taken $\delta=\eta$ to keep the coefficients of $(p_x^2+p_y^2)$ and $(x^2+y^2)$ in similar form as  appear in  Hamiltonian  (\ref{QH}) for the corresponding terms.  The above deformed Hamiltonian incorporates two additional terms   as momentum coupling term and angular momentum about $z$-axis with Kinetic and potential energies. This fact reveals that at the scale of space-space and space-momentum noncommutativity the  momentum coupling terms and the angular momentum contribute  to the total energy with extra symmetries. 
\subsection{Momentum coupling term}
The above expression (\ref{NCH}) reveals  the dependency of energy  on the product of the two momenta, it is evident that the energy depends on $p_x p_y$ . This means that two displacements along orthogonal directions cannot be dynamically independent. So, $p_x p_y$ can be viewed as a correlation between motions along the $x$-axis and $y$-axis. The sign of their product is negative, i.e., $p_x p_y < 0$ when the two momenta have opposite signs, leading to negative contributions in the total energy. We can further define  $p_+=\frac{p_x+p_y}{\sqrt2}, \qquad p_-=\frac{p_x-p_y}{\sqrt2}$. Thus, we have $p_xp_y=\frac14\left[(p_x+p_y)^2-(p_x-p_y)^2\right]$. It turns out that the interaction term $p_x p_y$ is a difference of kinetic energies along two rotated coordinates and implies the kinetic-energy mixing of two oscillators.

\subsection{Rotational part of Hamiltonian: Angular momentum}
Let us assume that we have the following relation $L_z|n,m\rangle=m\hbar|n,m\rangle$for the state of the oscillator and let's consider what would happen with the application of our Hamiltonian on it as $H_L|n,m\rangle = \frac{M\Omega^2\epsilon}{2\hbar} (m\hbar)|n,m\rangle$. Therefore, energy shift will be $\Delta E_m= \frac12M\Omega^2\epsilon\,m$ . In turn, oscillator states characterized by angular momenta$+m\hbar \quad\text{and}\quad -m\hbar$ will experience different shifts of their energies. The coupling of $L_z$ leads to the degeneracy removal between the states with clockwise and counterclockwise motion. This is because there is a $+\frac{M\Omega^2\epsilon}{2\hbar}L_z$ term in the Hamiltonian which gives the mathematical expression for a case where there exists an effective rotational/angular-momentum coupling for the oscillator. This enables the definition of an effective angular velocity-like parameter which is $\omega_{\rm eff} = -\frac{M\Omega^2\epsilon}{2\hbar}$ and now $H_L=-\omega_{\rm eff}L_z$. This means that the oscillator is not literally rotating about the z axis but rather that there exists an effective angular momentum coupling.

These extra terms are the physical signatures of the noncommutative structure. As transforming the noncommutative variables into ordinary canonical variables, the extra terms  vanish and we reattain the usual Hamiltonian for the Quantum Harmonic oscillator 
\section{The equation of motion in Noncommutative Variables}
This section includes the derivation of the  equation of motion for the  $2$-D NC harmonic oscillator in terms of NC parameters  directly from Hamilton's equations. Finally that equation of motion is reduced to NC coordinates by using transformations  (\ref{Tr}).
\subsection{ The equation of motion in terms of NC parameters}
The equation of motion associated with $2$-D NC Harmonic oscillator can be calculated as below by using the Hamilton's equations as below
\begin{equation}\label{2D}
\left\{
\begin{aligned}
\ddot x &= - \omega^2 x-  \frac{\epsilon \omega^2}{\hbar}p_y - \frac{\epsilon}{\hbar} \ddot p_y
 \\
\ddot y &= - \omega^2 y+  \frac{\epsilon \omega^2}{\hbar}p_x + \frac{\epsilon}{\hbar} \ddot p_x
\end{aligned}
\right..
\end{equation}
 This can be shown that the trivial solutions $x=y= e^{i\omega t}$, $p_y=p_y=  e^{ \pm i\omega t}$ satisfy above equation of motions. Further,  under the commutative limit $\theta^{\mu \nu} \longrightarrow 0$ for all values of $\mu$ and $\nu$, the noncommunicative terms vanish and we obtain homogeneous equation of motion for the $2$-D classical Harmonic oscillator with  total energy  becomes $ H =  \frac{1}{2m}  p_x^2 
+ \frac{1}{2m} p_y^2 
+ \frac{m \omega^2}{2} x^2 
+  \frac{m \omega^2}{2} y^2$ .
\subsection{Reduction to NC coordinates $(X,Y)$}
The above equation of motion (\ref{2D}) for  $2$-D NC Harmonic oscillator can be expressed in terms of Purely NC space coordinates by using the transformations  (\ref{Tr}) as below
\begin{equation}\label{2D1}
\left\{
\begin{aligned}
\ddot X &= - \omega^2 X
 \\
\ddot Y &= - \omega^2 Y
\end{aligned}
\right.\end{equation}
These equations of motion are homogeneous  and representing $2$-D NC Harmonic oscillator oscillating with angular frequency $\omega$, the unperturbed actual frequency.

\section{Super integrability in noncommutative framework}
Let us define the  mixed tensor \( Q_{jk} \) as below
\begin{equation}
Q_{jk} = \frac{1}{2m} P_j P_k + \frac{1}{2} m \omega^2 X_j X_k.
\end{equation}
Explicitly we can construct three tensors   in terms of old phase coordinates and noncommmutative parameters as below
\begin{align}
Q_{11} &= \frac{p_x^2}{2 m} + \frac{1}{2} m \omega^2 x^2 + \frac{p_y^2 \eta^2}{2 m \hbar^2} + \frac{m p_y^2 \epsilon^2 \omega^2}{2 \hbar^2} + \frac{p_x p_y \eta}{m \hbar} + \frac{m p_y x \epsilon \omega^2}{\hbar}, \\[2mm]
Q_{22} &= \frac{p_y^2}{2 m} + \frac{1}{2} m \omega^2 y^2 + \frac{p_x^2 \delta^2}{2 m \hbar^2} + \frac{m p_x^2 \epsilon^2 \omega^2}{2 \hbar^2} + \frac{p_x p_y \delta}{m \hbar} - \frac{m p_x y \epsilon \omega^2}{\hbar}, \\[1mm]
Q_{12} &= \frac{p_x p_y}{2 m} + \frac{1}{2} m \omega^2 x y + \frac{p_x p_y \delta \eta}{2 m \hbar^2} - \frac{m p_x p_y \epsilon^2 \omega^2}{2 \hbar^2} + \frac{p_x^2 \delta}{2 m \hbar} + \frac{p_y^2 \eta}{2 m \hbar} - \frac{m p_x x \epsilon \omega^2}{2 \hbar} + \frac{m p_y y \epsilon \omega^2}{2 \hbar}.
\end{align}
The angular momentum in terms of noncommutative parameters can be calculated as 
\begin{equation}
L_z= xp_y -yp_x  + \frac{\delta}{ \hbar} (xp_x -yp_y) + \frac{\epsilon }{\hbar} (p^2_x + p^2_y) + 2 \frac{\epsilon  \delta}{\hbar^2} p_x p_y.
\end{equation} This can be shown the Moyal bracket of each component of  mixed tensor $Q_{jk}$ with Hamiltonian( \ref{NCH}) vanishes as $[H, Q_{11}] = 0, [H, Q_{22}] = 0  [H, Q_{12}] = 0$  which ensures  the super integrability of $2$ -D Harmonic oscillator  in noncommutative settings.
\section{NC Hamiltonian in terms of creation and annihilation operators}
At this stage, we can easily define the creation and annihilation operator in terms of modified mass $M$ and angular velocity $\Omega$ as below ,
\begin{equation}
\left\{
\begin{aligned}
a_{x_{j}} &= \sqrt{\frac{M\Omega}{2\hbar}} \,  \left( {x_{j}} + \frac{i p_{x_{j}}}{M \Omega}\right), 
&\quad a_{x_{j}}^\dagger &= \sqrt{\frac{M\Omega}{2\hbar}}  \, \left({x_{j}} - \frac{i p_{x_{j}}}{M \Omega}\right), \\
{x_{j}}   &= \sqrt{\frac{\hbar }{2 M\Omega}} \left(a_{x_{j}}^\dagger + a_{x_{j}}\right), 
&\quad p_{x_{j}} &= i \sqrt{\frac{M\Omega \hbar}{2}} \left(a_{x_{j}}^\dagger - a_{x_{j}}\right). \\
\end{aligned}
\right.
\end{equation}
In above expressions $j=1, 2$ and $x_1= x, x_2=y$, now the Hamiltonian (\ref{NCH})can easily  be transformed into  annihilation and creation operators as below  
\begin{align}\label{NCH1}
H &= (a_x^\dagger a_x + a_y^\dagger a_y + 1 )\Omega \hbar
- 2f G +
g  \left(  a_y^\dagger a_x   - a_x^\dagger a_y  \right)= H^\dagger .
\end{align}

Here $f=M\delta  \Omega D B $,  $g=\frac{i \epsilon M \Omega^2 }{4}$ Above resulting expression involves number operators and and mixed creations, annihilation operators with $G=a_x^\dagger a_y^\dagger  - a_x^\dagger a_y - a_x a_y^\dagger + a_x a_y $.  This also can be shown that the Hamiltonian is hermitian in noncommuative settings.
\section*{Remark $5.1$} Under  the commutative limit $\theta^{\mu \nu} \longrightarrow 0$ for all values of $\mu$ and $\nu$, we obtain original Hamiltonian $ H = (a_x^\dagger a_x + a_y^\dagger a_y + 1 )\Omega \hbar$ as for the quantum harmonic oscillator. This is obvious that Hamiltoinain  remains  hermitian and here $r= \frac{\epsilon M \Omega^2 }{4}$, $s=2M\delta  \Omega$. The noncommutativity preserves the hermiticity of the Hamiltonian, it means the eigenvalues of Hamiltonian for $2$-D NC Harmonic oscillator with noncommuting canonical coordinates   are reals. 
 \section{Expectation Value of the Hamiltonian}
Let we define the number operators as $N_1 = a_x^\dagger a_x, N_2 = a_y^\dagger a_y$.
These operators act on the state $\lvert n_x, n_y \rangle$ 
\begin{equation}
\begin{cases}
N_1 \lvert n_x, n_y \rangle = n_1 \lvert n_x, n_y \rangle, \\
N_2 \lvert n_x, n_y \rangle = n_2 \lvert n_x, n_y \rangle.
\end{cases}
\end{equation}
here the  $n_1$ and $n_2$ are the eigenvalues of operators $N_1$ and $N_2$ representing the number of particles along $x$-axis and $y$-axis respectively. Now we may propose some more operators involving creation and annihilation along different axis  as below 
\begin{equation}
\begin{cases}
N_{12} = a_x a_y, \quad N_{12}^\dagger = a_y^\dagger a_x^\dagger , \\
M_{12} = a_x^\dagger a_y, \quad M_{12}^\dagger = a_y^\dagger a_x.
\end{cases}
\end{equation}
The action of thees mixed operators on the state $\lvert n_x, n_y \rangle$ are defined as 

\begin{equation} 
\begin{cases}
N_{12} \lvert n_x, n_y \rangle =\sqrt{(n_1)(n_2)}\lvert n_x -1, n_y-1 \rangle  , \\
M_{12} \lvert n_x, n_y \rangle = \sqrt{(n_1+1)(n_2+1)}\lvert n_x +1, n_y+1 \rangle.
\end{cases}
\end{equation}
Then the action of Hamiltonian on state   $\lvert n_x, n_y \rangle$ can be calculated as

\begin{equation}
\begin{aligned}
H|n_x,n_y\rangle
=&\;
\hbar\Omega (n_x+n_y+1)|n_x,n_y\rangle
\\[2mm]
&-2f\sqrt{(n_x+1)(n_y+1)}
\,|n_x+1,n_y+1\rangle
\\[2mm]
&+(2f-g)\sqrt{(n_x+1)n_y}
\,|n_x+1,n_y-1\rangle
\\[2mm]
&+(2f+g)\sqrt{n_x(n_y+1)}
\,|n_x-1,n_y+1\rangle
\\[2mm]
&-2f\sqrt{n_xn_y}
\,|n_x-1,n_y-1\rangle .
\end{aligned}
\end{equation}\label{NCH4}
The matrix associated with  Hamiltonian (\ref{NCH4}) includes non-zero elements at off diagonal positions due to the extra coupling terms emerge with noncommutative structure. The true energy eigenvalues are obtained by diagonalizing the
Hamiltonian matrix in the $\{|n_x,n_y\rangle\}$ basis or by
performing a suitable Bogoliubov (canonical) transformation to
new bosonic operators that diagonalize $H$.
\section*{Eigenvalues}
The number state $|n_x,n_y\rangle$ is an eigenstate only if the
off-diagonal couplings vanish (for example, $f=0$ and $g=0$).
In that case,

\begin{equation}
\boxed{
E_{n_x,n_y}
=
\hbar\Omega\,(n_x+n_y+1)
}
\end{equation}
For $f\neq0$ or $g\neq0$, the Hamiltonian mixes different number
states, so $|n_x,n_y\rangle$ is not an eigenstate.
\section{Noncommutative Painlev\'e second equation}
Here we derive a NC analog of Painlevé second equation \cite{PPE} with the presentation of new phase coordinates. These transformations connect the NC phase coordinates $(Q,P)$ to the canonical coordinates of Okamoto Painlev\'e second Hamltonian \cite{OKA}. The quantum and NC versions of that equation have been derived in different frameworks. For example,  quantum Painlev\'e second has been derived in  \cite{HG1, HG2} by  Heisenberg quantum commutation approach and subsequently its NC anlog is presented by V. Rtakh and V. Rubtsov \cite{VV} through the  NC Toda lattice. Later its quansideterminant solutions with NC expression were derived by Irfan  \cite{IM1, IM2} through its Lax representation with the implementation of Darboux transformation.  Recently, the integrability of three matrix versions of the Painlev\'e II equation have be discussed in \cite{AD} in framework of Painlev\'e–Kovalevskaya test. In this context with growing interests in theory of Painlev\'e equations is  strengthened with more new results as presented in \cite{PAD1, PAD2} on Painlev\'e second equation and its partner equation for their matrix cases. \\
\section*{Proposition $ 7.1$}
The following Hamiltonian 
\begin{equation} \label{NCPH}
H(Q,P) 
= \frac{1}{2} P^2 + QPQ +\frac{z}{2} P -(\alpha_2 - \frac{1}{2})Q.
\end{equation}
in terms of NC phase coordinates $(Q,P)$ yields NC  Painlev\'e second equation  by using Hamilton's equations.
\section*{Proof:}
By using the Hamilton's equations one may calculate the first order differential equation as
 $Q^{'} = P+Q^2 + \frac{z}{2}$ and $P^{'} = -PQ-QP + \alpha_2 - \frac{1}{2}$ from Hamiltonian (\ref{NCPH}). Now after eliminating $P$ from these first order differential equation we obtain NC painlev\'e second equation for $Q$ as below
\begin{equation}\label{NCP1}
Q^{''} = 2Q^3 + \frac{1}{2} [z,Q]_{+} + \alpha_2. 
\end{equation}
Now from Moyal product (\ref{1}) this can be shown that $[Q,P]_{-} = i \theta^{12}$, here  $\theta^{12}$ is anti-symmetric parameter defines the noncommutativity. We may introduce the following transformations
\begin{equation}\label{Tr2}
\left\{
\begin{aligned}
& Q = q + \frac{\theta^{12}}{\hbar}z, \\
& P= -p + \frac{\theta^{21}}{\hbar} z,
\end{aligned}
\right.
\end{equation}
here $(Q,P)$ are purely  noncommuting phase coordinates and $ (q,p)$ are the canonical coordinates which satisfy $ q^{''} = 2q^3 - zq + \alpha_1 $ and $p^{'} = 2pq - \alpha_1 + \frac{1}{2}$. As in case of Heisenberg canonical  quantization we have $[z,p]_{-} = i\hbar $ and with this second equation in (\ref{Tr2} )
 gives $[z,P]_{-}= i\hbar$. For $z$ and $Q$ relation we recall $P^{'} = PQ+QP - \alpha_2 + \frac{1}{2}$  for  calculating $[z,Q]_{-} $ which yields $ P[z,Q]_{-}+ [z,Q]_{-} P=-2i \hbar Q$.
In above transformation (\ref{Tr2}), $\theta^{i j}$ are the components of following anti-symmetric tensor
\begin{align}
 \theta^{\mu \nu}=\left(\begin{array}{cc}
0 & \epsilon  \\
-\epsilon & 0 
\end{array}\right), 
\end{align}\
 The transformation (\ref{Tr2}) yield $[Q,P]_{-} = i \epsilon$  which may be calculated directly from Moyal noncommutative product for arbitrary variables.  Now under the phase coordinates transformations (\ref{Tr2}) the above Hamiltonian (\ref{NCPH})  will take the following form
\begin{equation} \label{NCPH}
H(q,p; \alpha , \epsilon) 
= H(q,p; \alpha ) + \frac{\epsilon}{\hbar} (q-1)(pz+zp) + \frac{\epsilon}{\hbar}z (\frac{\epsilon}{2\hbar}z-\frac{\epsilon}{\hbar}q^2-2\frac{\epsilon}{\hbar}qz+ \frac{\epsilon}{\hbar}pz -\frac{\epsilon^2}{\hbar^2} z^2).
\end{equation}
Here in above expression  $H(q,p; \alpha ) =\frac{1}{2} p^2 - pq^2 - \frac{z}{2} p -(\alpha_1 + \frac{1}{2})q$  Okamoto Painlev\'e second Hamiltonian can be retained only if $\alpha_2 =  \alpha _1 +1$.
\section*{Remark $7.1$ } This remarks includes the derivation of Yablonskii Vorob’ev first polynomial \cite{YAB, VOR}   as the solution of Painlev\'e second equation  with the help of proposed phase space  coordinates transformation.   Let take the  derivation of  $Q^{'} = P+Q^2 - \frac{z}{2}$ and then using the transformation (\ref{Tr2}), we get $ Q^{'} +  q^{'}= -z^{-2}$. Now if we choose $Q= 0$ with  $\alpha_2 = 0$ for (\ref{NCP1}), these settings  together yield $q= z^{-1}$, her  $ \alpha _1 = -1$, which satisfies  original classical Painlev\'e second equation $ q^{''} = 2q^3 + zq + \alpha_1 $.  
\section*{Conclusion}
We have  discussed  the  application of Moyal  formalism which defines the  noncommutative  product for  any arbitrary functions or variables. We introduced  new  phase space coordinates transformation  which are  consistent  with the Moyal noncommunicative brackets.  These transformations are implied to construct the noncommutative analogs of $2$-dimensional QH oscillator  and  Painlev\'e second equation  through their associated noncommutative  Hamiltonians. The computations include a future motivation to propose phase coordinates transformations for $3$ -D quantum QH oscillator with noncommunicative analog for Hamilto Jacobi theory  and for the other Painlev\'e equations. In this context, future focus may involve the  generalization of proposed Painlev\'e second NC phase coordinates transformation  for  its partner equations and to  calculate their  polynomial solutions.
\section*{Acknowledgement}
Thankful to the University of the Punjab, Lahore and University of South Florida, USA for assisting to complete this project. 
\section*{Conflict of interest} No conflict of interest.
\section*{Funding} No any funding.


\begin{thebibliography}{99}

\bibitem{HJ}{ H.J. Groenewold, On the principles of elementary quantum mechanics, Physica 12 (1946) 405}
\bibitem{JE}{J.E. Moyal, Quantum mechanics as a statistical theory, Proc. Cambridge Philos. Soc. 45 (1949) 99.}
\bibitem{HT}{M. Hamanaka, K. Toda, J. Phys. A 36 (2003) 11981–11998, hep-th/0301213.}
\bibitem{HT1}{M. Hamanaka, K. Toda, Phys. Lett. A 316 (2003) 77–83, hep-th/0211148.}
\bibitem{MB}{ M.B. Sedra, Moyal noncommutative integrability and
the Burgers–KdV mapping, Nuclear Physics B 740 [PM] (2006) 243–270.}
\bibitem{NHO}{ B. Muthukumar and P. Mitra, Non-commutative oscillators and the
commutative limit, Phys. Rev. D 66 ( 2002) 027701 DOI: https://doi.org/10.1103/PhysRevD.66.027701}
\bibitem{HG1}{  H. Nagoya, Quantum Painlev\'e systems of type Al, Internat. J. Math. 15 (2004) 1007–1031, math.QA/0402281.}
\bibitem{HG2}{ H. Nagoya, B. Grammaticos, A. Ramani, Quantum Painlev\'e equations: from continuous to discrete, SIGMA 4 (051) (2008) 9 pages.} 
\bibitem{VV}{ Vladimir Retakh and Vladimir Rubtsov 2010 J. Phys. A: Math. Theor. 43 505204
DOI 10.1088/1751-8113/43/50/505204.}
\bibitem{WM}{W. X. Ma, An extended noncommutative KP hierarchy, Journal of Mathematical Physics, Vol.51(2010), 073505 }
\bibitem{IM1}{ Irfan Mahmood, Lax pair representation and Darboux transformation of noncommutative Painlev\'e’s second equation, Journal of Geometry and Physics,  62 (2012) 1575–1582.}
\bibitem{IM2}{ Irfan Mahmood, Quasideterminant solutions of NC Painlev\'e II equation with the Toda solution at n = 1 as a seed solution in its Darboux transformation, Journal of Geometry and Physics, 95 (2015) 127–136.}
\bibitem{AD}{Adler, V.E., Sokolov, V.V. Matrix Painlev\'e II equations. Theor Math Phys 207, 560–571 (2021). https://doi.org/10.1134/S0040577921050020}
\bibitem{PAD1}{Bobrova, I. A., and V. V. Sokolov. "On matrix Painlevé-4 equations." Nonlinearity 35.12 (2022): 6528-6556.}
\bibitem{PAD2}{Pickering, Andrew. On the integrability of a new multiparameter matrix second Painlev\'e equation. Chaos, Solitons and  Fractals 212 (2026): 119008.}
\bibitem{PPE}{P.Painlev\'e, Sur les \'equations diff\'erentiel les du second ordre et d’ordre superieur dont l’int\'egrale g\'en\'erale est uniforme, Acta. Math.25(1902)1–86.}
\bibitem{OKA}{Okamoto, K., Polynomial Hamiltonians associated with Painlev\'e equations, I, Proc. Japan
Acad. Ser. A Math. Sci., 56 (1980), 264-268; IL ibid, 56 (1980), 367-371.}
\bibitem{YAB}{A. L. Yablonskii, Vesti AN BSSR, ser. fiz-tekh, Nauk. 3, 30 (1959). 3) }
\bibitem{VOR}{ A. P. Vorob’ev, Differentsial’nye Uravneniya, Differen Equat. 1, 79
(1965). }

\end{thebibliography}
\end{document}